\documentclass[fleqn,usenatbib]{mnras}

\usepackage{newtxtext,newtxmath}
\usepackage[T1]{fontenc}
\usepackage{ae,aecompl}

\usepackage{graphicx}
\usepackage{amsmath}
\usepackage{amssymb}
\usepackage{enumerate}
\usepackage{ulem}
\usepackage{bm}
\usepackage{xcolor}
\usepackage{booktabs}

\title{Equation of state sensitivities when inferring neutron star and dense matter properties}

\author[S.K. Greif, G. Raaijmakers et al.]{
S.K. Greif,$^{1,2}$\thanks{Joint lead authors: sgreif@theorie.ikp.physik.tu-darmstadt.de, g.raaijmakers@uva.nl}
G. Raaijmakers,$^{3}${\color{blue}\footnotemark[1]}
K. Hebeler,$^{1,2}$
A. Schwenk,$^{1,2,4}$
A.L. Watts$^{3}$
\\
$^{1}$Institut f\"ur Kernphysik, Technische Universit\"at Darmstadt, D-64289 Darmstadt, Germany\\
$^{2}$ExtreMe Matter Institute EMMI, GSI Helmholtzzentrum f\"ur Schwerionenforschung~GmbH, D-64291 Darmstadt, Germany\\
$^{3}$Anton Pannekoek Institute for Astronomy, University of Amsterdam, PO Box 94249, 1090GE Amsterdam, the Netherlands\\
$^{4}$Max-Planck-Institut f\"ur Kernphysik, Saupfercheckweg 1, D-69117 Heidelberg, Germany
}

\date{Accepted XXX. Received YYY; in original form ZZZ}

\pubyear{2018}

\begin{document}
\label{firstpage}
\pagerange{\pageref{firstpage}--\pageref{lastpage}}
\maketitle

\begin{abstract}
Understanding the dense matter equation of state at extreme conditions is an important open problem. Astrophysical observations of neutron stars promise to solve this, with NICER poised to make precision measurements of mass and radius for several stars using the waveform modelling technique. What has been less clear, however, is how these mass-radius measurements might translate into equation of state constraints and what are the associated equation of state sensitivities. We use Bayesian inference to explore and contrast the constraints that would result from different choices for the equation of state parametrization; comparing the well-established piecewise polytropic parametrization to one based on physically motivated assumptions for the speed of sound in dense matter. We also compare the constraints resulting from Bayesian inference to those from simple compatibility cuts. We find that the choice of equation of state parametrization and particularly its prior assumptions can have a significant effect on the inferred global mass-radius relation and the equation of state constraints. Our results point to important sensitivities when inferring neutron star and dense matter properties. This applies also to inferences from gravitational wave observations.
\end{abstract}


\section{Introduction}

With initial results from the {\it Neutron Star Interior Composition Explorer} (NICER) \citep{NICER,NICER2} mission imminent, X-ray pulsar waveform modelling is poised to deliver its first precision constraints on the dense matter equation of state (EOS).  Emission from a hotspot such as the hot polar cap of an X-ray pulsar gives rise to a pulsation in the light emitted from the star as it rotates.  The shape and energy dependence of the waveform depend on the relativistic gravitational properties of the star because the light must propagate through the neutron star space-time; this in turn depends on the EOS of supranuclear density matter within the neutron star. Given good relativistic ray-tracing models for rapidly rotating neutron star space-times, it is possible to perform Bayesian inference of EOS parameters using the pulse profile data \citep[see][and references therein]{Watts16}. The waveform modelling technique is also central to plans to put even tighter constraints on the dense matter EOS with the next generation of large-area X-ray telescopes.  Two such concepts, the {\it enhanced X-ray Timing and Polarimetry mission} \citep[eXTP,][]{eXTP19,Watts19} and the {\it Spectroscopic Time-Resolving Observatory for Broadband Energy X-rays} \citep[STROBE-X,][]{STROBEX2}, are currently in development. 

One interesting question that arises is how to parametrize the EOS.  The form of the parametrization chosen has consequences for how one does the inference \citep{Riley18, Raaijmakers18} and one must also consider which parametrizations deliver most information about the microphysics of the particle interactions that are the ultimate goal. One very well-established model is the piecewise polytropic (PP) model \citep{Read09}, which was used to put general constraints on the EOS based on nuclear physics and observations \citep{Hebeler10,Hebeler13}. However, the PP construction implies discontinuities for other properties, such as the speed of sound. The speed of sound is also physically constrained by both causality requirements and an asymptotic limit at high density. This has led some, most recently \cite{Tews18}, to consider a sound-speed based parametrization.

In this paper, we develop the speed of sound (CS) parametrization further, and examine how it compares to the PP parametrization. We do this by exploring the scenarios expected to be delivered by NICER.  NICER has four primary targets, for which only one has a known mass.  We explore the types of constraints that might arise from both PP and CS parametrizations, depending on where in parameter space the sources of unknown mass end up being located. Our analysis also lets us explore how sensitive the inference of neutron star and dense matter properties is to both the choice of EOS parametrization and the priors that are specified for the parameters of a given model. This is of relevance to all efforts to infer neutron star and EOS properties from observational data, including gravitational wave observations of neutron star mergers like GW170817 \citep{Abbott17}, see, e.g., \cite{Abbott18,Annala18,Most18,Tews18b,Lim18tidaldef} that use parametrized EOS models.

\section{Speed of sound parametrization}
\label{sec:CS_parametrization}

The most commonly used EOS parametrizations in the literature, the PP model \citep{Read09} and the spectral model \citep{Lindblom10}, parametrize the EOS space directly. Here we will in addition choose to parametrize the speed of sound in the neutron star interior and compare it to the established PP parametrization. To determine the functional form of such a parametrization we take into account constraints on the speed of sound coming from both theoretical calculations and observations. First, the speed of sound in the asymptotic high-density limit should converge to $c_{s}/c = 1/\sqrt{3}$ according to calculations in perturbative quantum chromodynamics (pQCD) \citep{Frag14pQCD}. These calculations have furthermore shown that the speed of sound should converge to this value from below, i.e. the leading-order corrections to this limiting value are negative. The measurement of two-solar-mass neutron stars \citep{Demorest10, Antoniadis13}\footnote{The mass of PSR J1614-2230 was recently updated from $1.97 \pm 0.04 \, M_\odot$ \citep{Demorest10} to $1.928 \pm 0.017 \, M_\odot$ \citep{Fonseca2016}.} on the other hand suggests that in the density regime between nuclear saturation density and the high-density limit the speed of sound should exceed the value of $1/\sqrt{3}$. In particular, \citet{Bedaque15} have shown that no EOS can produce such a high neutron star mass unless this value is exceeded in some density regime. That means the CS parametrization needs to allow for an increase of the speed of sound beyond the limit $c_{s}/c = 1/\sqrt{3}$ at intermediate densities.

In order to describe the asymptotic behaviour of the speed of sound we employ a logistic function, which at low densities is matched to the theoretical calculations of neutron star matter performed in \citet{Hebe10nmatt,Hebeler13} using nuclear interactions based on chiral effective field theory (cEFT) \citep{Epel09RMP,Hamm13RMP}. The results of these microscopic calculations are used up to a density of $n_{\mathrm{cEFT}} = 1.1 n_0$, with the nuclear saturation density $n_0 = 0.16 \: \text{fm}^{-3}$. For the density region $n \le 0.5 n_0$ the BPS crust EOS is used \citep{Baym71, Negele73}. Remarkably, at the transition density $n = 0.5 n_0$ the results for the EOS of both approaches are consistent with each other (see \citealt{Hebeler13} for details).

\begin{figure}
\centering
\includegraphics[width=.9\columnwidth]{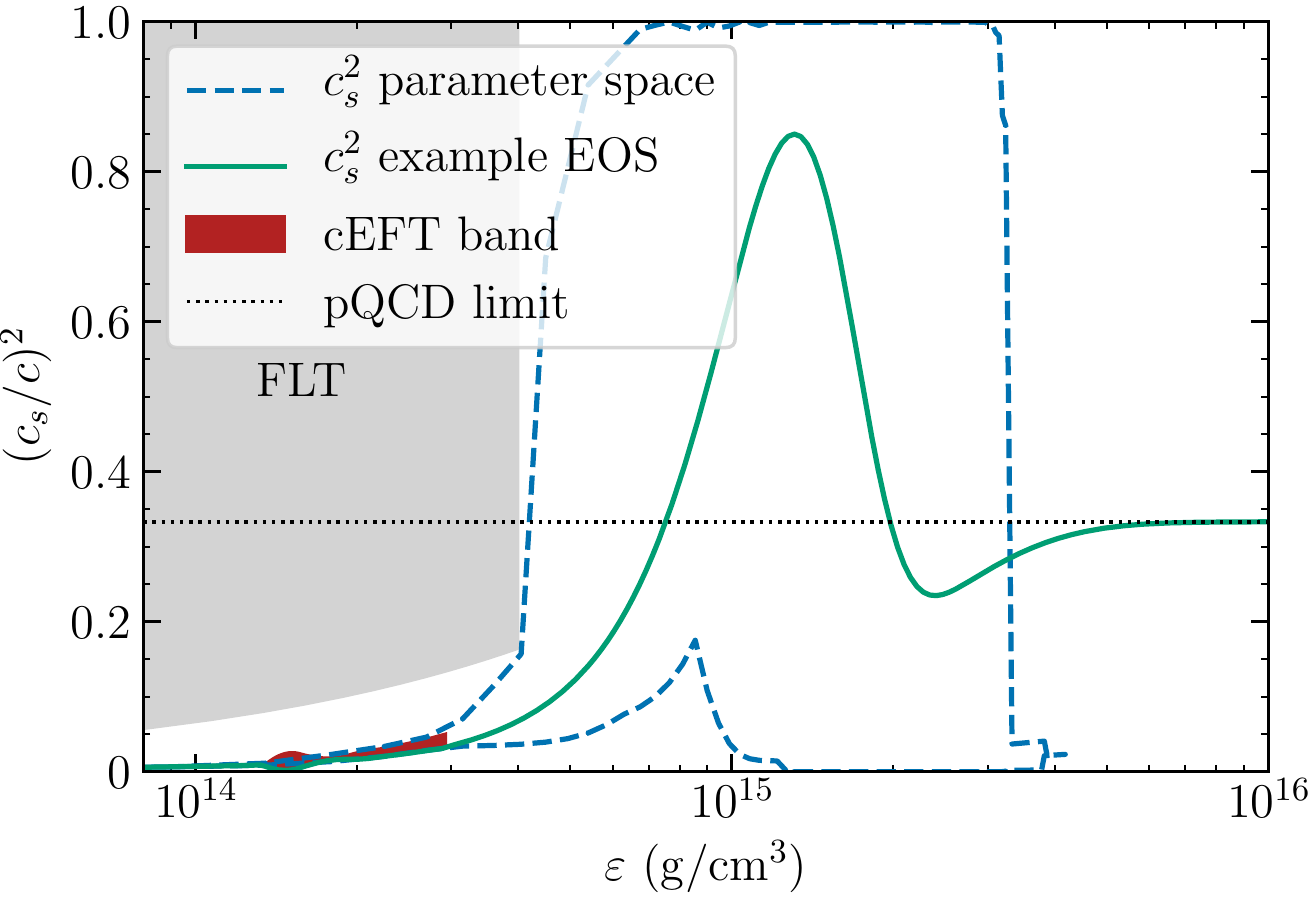}
\caption{Illustration of the CS parametrization given in Eq.~(\ref{eq:cS_parametrization}). The square of the speed of sound is represented by the solid green line. At the transition density $\varepsilon_{\mathrm{cEFT}}$ the model is matched to the cEFT band, while for very high densities the speed of sound converges from below to the limit of $1/\sqrt{3}$ that is shown by the black dotted line. The blue dashed outline represents the parameter space that is consistent with the constraints listed in the text. The grey area at low densities represents the excluded region by the Fermi liquid theory (FLT) constraints. Details are given in the text.}
\label{fig:fig1}
\end{figure}

\begin{figure*}
\centering
\includegraphics[width=0.9\textwidth]{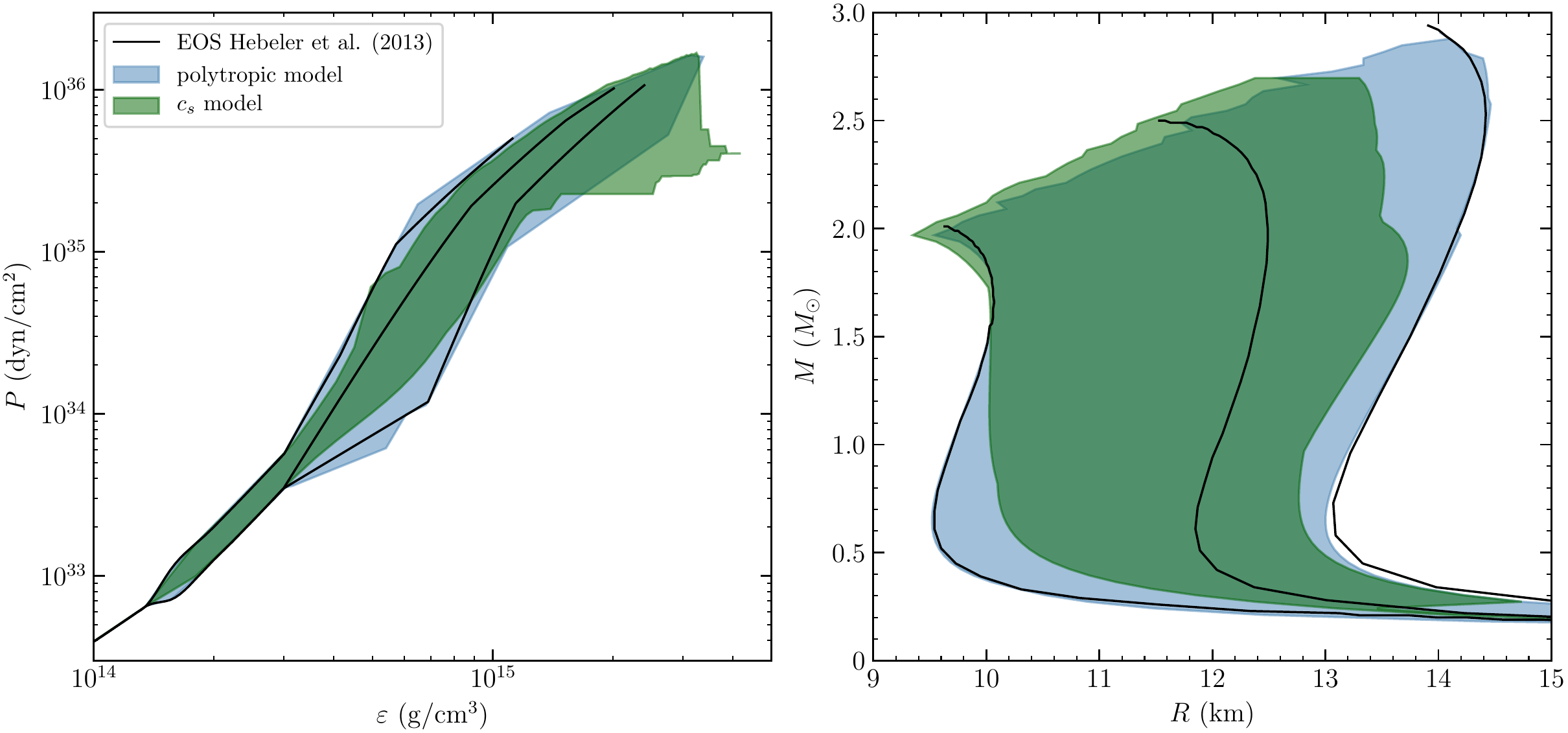}
\caption{Comparison of the ranges for the EOS (left panel) and mass-radius relations of neutron stars (right panel) based on the PP and CS parametrizations after including all constraints discussed in Section~\ref{sec:parameterspace}. The black lines show the three representative EOS of the PP model \citep{Hebeler13}, the light blue bands show the uncertainty ranges resulting from the PP model and the darker green bands those from the CS model.}
\label{fig:fig2}
\end{figure*}

The increase of the speed of sound at intermediate densities is modeled by a Gaussian function. The complete CS parametrization can then be written as 
\begin{equation}
c_{s}^{2}(x)/c^2 = a_1 \, e^{-\frac{1}{2}\left( x - a_2\right)^2 /a_3^2} + a_6 + \frac{\frac{1}{3} - a_6}{1 + e^{-a_5 (x - a_4)}} \,,
\label{eq:cS_parametrization}
\end{equation}
with $x\equiv \varepsilon/(m_{\mathrm{N}} n_0)$ and the nucleon mass \mbox{$m_{\mathrm{N}} = 939.565 \, \mathrm{MeV}$}. The parameter $a_6$ is fitted to match to the upper and lower limit of the band predicted by cEFT at the transition energy density $\varepsilon_{\mathrm{cEFT}} = \varepsilon (n_{\mathrm{cEFT}})$\footnote{Note that the transition densities $\varepsilon_{\mathrm{cEFT}}$ differ for the lower (min) and upper (max) limit of the cEFT band, with $\varepsilon_{\mathrm{cEFT}}^{\mathrm{min}} = 167.8 \, \mathrm{MeV} \, \mathrm{fm}^{-3}$ and $\varepsilon_{\mathrm{cEFT}}^{\mathrm{max}} = 168.5 \, \mathrm{MeV} \, \mathrm{fm}^{-3}$ \citep{Hebeler13}.}. One could in principle match continuously to the cEFT band, but this requires information on the functional form of the EOS within the band (in order to obtain the speed of sound). Because the mass and radius of a neutron star are only weakly dependent on the EOS at these densities, we expect that our main findings will however not depend on this particular choice. The values of the other free parameters, $a_1$ to $a_5$, are allowed to vary freely over a wide range of values, and are only limited by physical constraints which are discussed in more detail in the next section. The parametrization (\ref{eq:cS_parametrization}) allows us to generate a large range of different types of EOSs, ranging from very soft to very stiff. The qualitative form of the speed of sound resulting from this parametrization is shown in Fig.~\ref{fig:fig1}. 

Starting from the speed of sound, the pressure as a function of energy density is then obtained via integration,
\begin{align}
P(\varepsilon) = \int_{0}^{\varepsilon} d \varepsilon^{\prime} \, c_{s}^{2}(\varepsilon^{\prime}) / c^2 \,.
\end{align}
The form of our CS parametrization is very similar to the one used by \citet{Tews18}. The two major differences are their use of an exponential function to model the asymptotic behaviour of $c_{s}$ at large densities as compared to our logistic function; and their use of a skewed Gaussian, where we use a non-skewed Gaussian. We have performed additional computations with a model that include a skewed Gaussian and found the results presented in this paper to be robust. In addition to these differences in the parametrization, we also apply different constraints in the low-density region. In particular, \citet{Tews18} used a less conservative value for the upper density limit $n_{\mathrm{cEFT}} = 2.0 \, n_0$ (rather than $n_{\mathrm{cEFT}} = 1.1 \, n_0$ in this work), which leads to tighter constraints for the EOS. The precise value of the breakdown density scale for nuclear interactions in nuclear matter is still an open question and subject to current research. Note that there are certainly other choices for the functional form of a CS-based parametrization that would meet the physical criteria we have used in its formulation; we are using this particular example here to illustrate the effects of different choices for the EOS model.

\begin{figure*}
\centering
\includegraphics[width=0.9\textwidth]{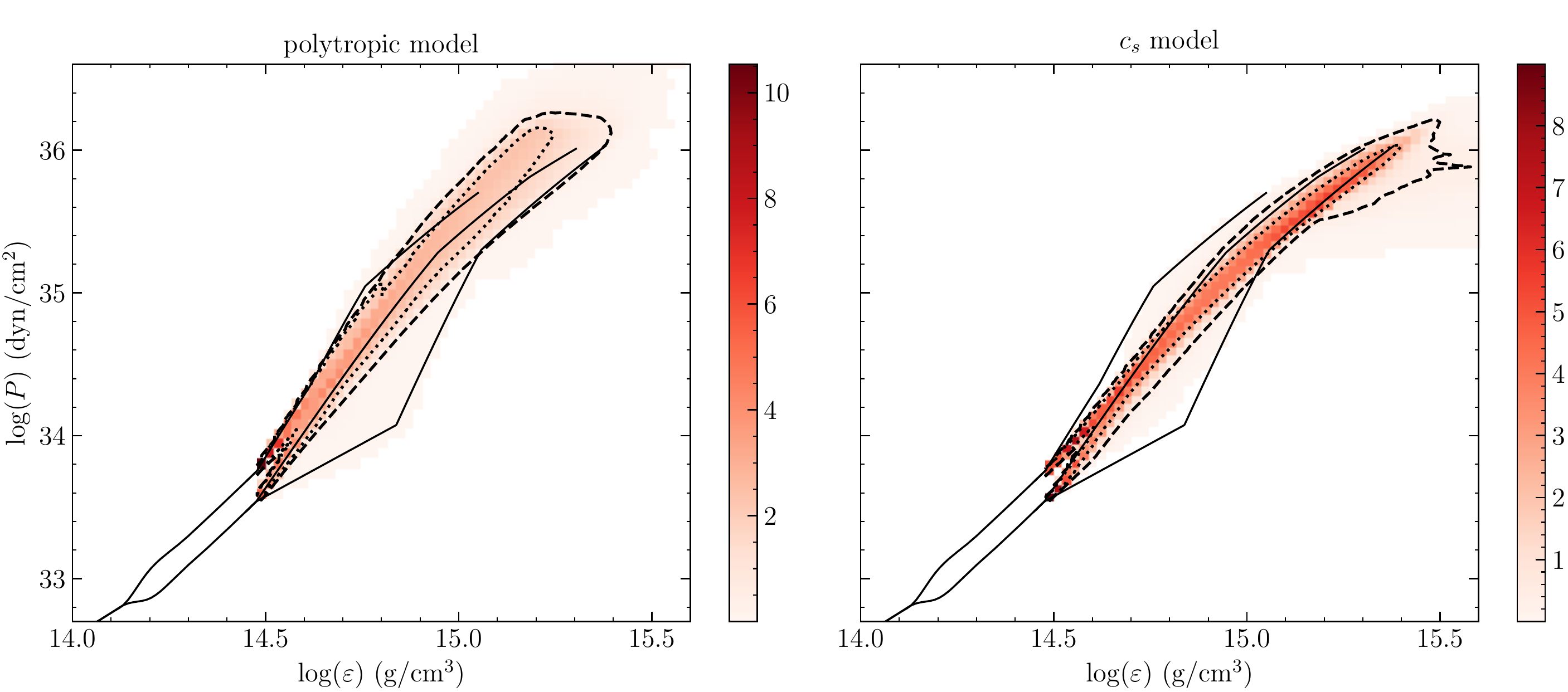}
\includegraphics[width=0.9\textwidth]{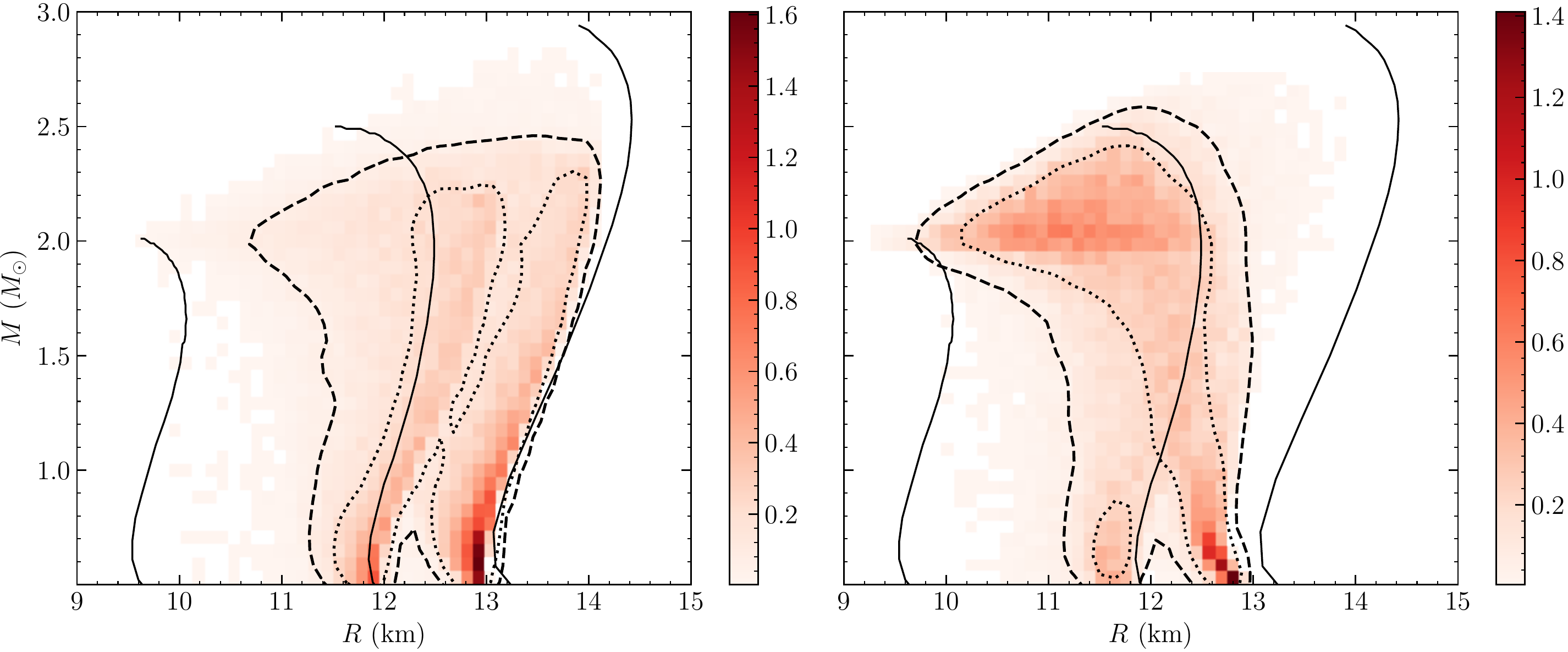}
\caption{\textit{Top:} Prior probability distributions transformed to the space of pressure as a function of energy density for the PP (left) and CS parametrization (right panel). The dotted and dashed lines indicate the 68\% and 95\% credible regions of the distributions, while the solid lines are the representative EOS from \citet{Hebeler13}. Both prior distributions exhibit a narrow region where most of the probability density is clustered, which falls off steeply towards higher and lower pressures. \textit{Bottom:} Similar to the upper panels, but now the prior distributions are transformed to the space of mass and radius. We observe that the prior constraints from Section~\ref{sec:parameterspace} result in a higher probability density towards larger radii for both parametrizations. The bimodal feature of the distributions are caused by the way the models have been matched to the lower and upper limit of the cEFT band.}
\label{fig:fig3}
\end{figure*}

\subsection{Definition of the parameter space}
\label{sec:parameterspace}

\subsubsection{Speed of sound parametrization}
\label{sec:CS_parameterspace}

For the generation of individual EOS based on the CS parametrization (\ref{eq:cS_parametrization}) we allow the free parameters to vary over a wide range of values and retain only those EOS that fulfil the following physical constraints:
\begin{enumerate}[(i)]
\item Each EOS is required to be able to support at least a $1.97 \ M_{\odot}$ neutron star, which is the lower $1\sigma$ limit of the heaviest precisely measured neutron star PSR J0348+0432 \citep{Antoniadis13}.
\item The speed of sound for each EOS must be causal, i.e., be lower than the speed of light, for all energy densities relevant in neutron stars. If the EOS becomes acausal before the maximum mass is reached, we discard these parameter values.
\item The speed of sound for each EOS must converge to $1/\sqrt{3}$ from below at least for asymptotically high densities ($\sim 50 \ n_0$) as determined by pQCD calculations \citep{Frag14pQCD}.
\item Whenever the speed of sound for an EOS is negative, we set $c_s = 0$. This allows for regions of constant pressure as would be the case for a first-order phase transition. 
\item For densities $n \le 1.5 \, n_0$ we assume that the bulk properties of matter can be described as a normal Fermi liquid. In Landau Fermi liquid theory (FLT) \citep{BaymPethick}, the speed of sound is given by:
\begin{equation}
c_{s, \text{FLT}}^{2}(n)/c^2 = \frac{1 + F_0}{m_{\mathrm{N}}^*/m_{\mathrm{N}}} \frac{1}{3 m_{\mathrm{N}}^2} \left( 3 \pi^2 n \right)^{\frac{2}{3}} \,,
\end{equation}
where $F_0$ denotes the spin-independent and isotropic ($l=0$) Landau parameter characterizing particle interactions. In FLT, nucleons are described in terms of effective degrees of freedom, so-called quasiparticles, with effective mass $m_{\mathrm{N}}^*$. The dimensionless Landau parameter $F_0$ is expected to be attractive, and calculations for neutron matter suggest $F_0 \approx -0.5(2)$ as well as $m_{\mathrm{N}}^*/m_{\mathrm{N}} \approx 0.9(2)$ at saturation density \citep{Schw02neutmat,Schw03noncentral}. Moreover, both $1+F_0$ and $m_{\mathrm{N}}^*/m_{\mathrm{N}}$ are of expected to be of order one. Given the above considerations, it is very conservative to assume $\frac{1 + F_0}{m_{\mathrm{N}}^*/m_{\mathrm{N}}} \le 3$ up to $1.5 \, n_0$. This implies
\begin{equation}
c_{s, \text{FLT}}^{2}(n)/c^2 \le \frac{1}{m_{N}^2} \left( 3 \pi^2 n \right)^{\frac{2}{3}} \,,
\end{equation}
which amounts to $c_{s, \text{FLT}}^{2} \le 0.163 c^2$ for $n = 1.5 \, n_{0}$.
We discard any EOS that exceeds this value for $n \le 1.5 \, n_0$.
While this choice is very conservative, we note that it affects the specific upper radius limit of the resulting mass-radius uncertainty bands.
\end{enumerate}

For our practical calculations we choose the following ranges for the parameters in the CS parametrization~(\ref{eq:cS_parametrization}). While isolated parameters outside these bounds may exist that result in a stable EOS satisfying the above constraints, we have checked that pushing these bounds further does not significantly affect the ranges presented in Fig.~\ref{fig:fig2}:
\begin{enumerate}[(i)]
\item[(1)] For the normalization of the Gaussian, we require $0.1 \leq a_1 \leq 1.5$. Lower values do not produce an EOS that supports a $1.97 \, M_{\odot}$ neutron star, at least for our sample of $\sim 10^5$ EOS, and higher values result in an acausal EOS. 
\item[(2)] The bounds on the mean of the Gaussian are taken to be $1.5 \leq a_2 \leq 12$.
\item[(3)] For the width of the Gaussian in terms of the mean we take the following range $0.05 \leq a_3/a_2 \leq 2$. 
\item[(4)] The mean of the logistic function is taken to be within $1.5 \leq a_4 \leq 37$. 
\item[(5)] For the steepness of the logistic function we take $ 0.1 \leq a_5 \leq 1$. 
\end{enumerate}
As discussed above, the last parameter, $a_6$, is fixed by matching to either the upper or the lower limit of the cEFT band at the transition energy density $\varepsilon_{\mathrm{cEFT}}$. Within the bounds (1)--(5) for the five parameters mentioned above, we calculate on a grid a large sample of different EOS and retain only those that fulfill all of the constraints (i)--(v). This calculation is only used to give an indication of the possible band that the EOS in the CS model spans in Fig.~~\ref{fig:fig2}. During the Bayesian inference described in Section~\ref{sec:bayesianmethod}, all parameters in the model are considered continuously, subject to the prior bounds and constraints described above, so that for the Bayesian inference effectively an even larger sample is considered.

\subsubsection{Piecewise polytropic parametrization}
\label{sec:PP_parameterspace}

Next, we summarize for completeness and comparison the details and parameter space of the PP parametrization as used in \cite{Hebeler13}. The part of the EOS up to the transition density is the same as for the CS parametrization used in this work. It consists of the BPS crust EOS and the cEFT band up to $n_{\mathrm{cEFT}} = 1.1 \, n_0$. For higher densities the EOS is extended using piecewise polytropes with three segments. The parameter ranges for the polytropic exponents are $1.0 \le \Gamma_{1} \le 4.5$, $0 \le \Gamma_{2} \le 8.0$, and $0.5 \le \Gamma_{3} \le 8.0$. The values for $\Gamma_{i}$ are varied using a stepsize of 0.5. The densities between the polytropes are $1.5 \, n_{0} \le n_{12} < n_{23} < n_{\text{max}}$, where $n_{\text{max}} \approx 8.3 \, n_{0}$ is found to be the maximum central density reached. The transition densities $n_{ij}$ are varied in steps of $0.5 \, n_{0}$. In this approach only the neutron star mass constraint (i) and causality constraint (ii) of Section~\ref{sec:CS_parameterspace} are used. Instead of the FLT constraint (v), note that the range of the first polytropic index $\Gamma_1$, which also controls the stiffness in the density range $n_{\mathrm{cEFT}} \le n \le 1.5 \, n_0$, was restricted in \cite{Hebeler13} to a smaller range $1.0 \le \Gamma_{1} \le 4.5$.

\subsubsection{Comparison of the parametrizations}

In Fig.~\ref{fig:fig2}, we compare the EOS and mass-radius bands resulting from the PP and CS parametrization, after including all  constraints discussed above. In general, the PP model covers a larger area in EOS space as well as in mass-radius space. This is for the following reasons. First, the constraints at low densities are not exactly equivalent in both parametrizations. For the PP model the range of the polytropic index $\Gamma_1$ was restricted explicitly, whereas for the CS parametrization the FLT constraint was employed. It turns out that the upper FLT limit is more constraining, such that more stiff EOS and hence more neutron stars with larger radii are discarded by this constraint. This shows that the upper radius limit of a typical neutron star is quite sensitive to the particular choice of constraints at nuclear densities. In addition, because the PP model allows for strong stiffening after a first-order phase transition, an EOS can be very soft for small to intermediate densities and get very stiff at higher densities, such that the mass constraint is still fulfilled. While the CS model allows for phase transitions as well, it does not allow the EOS to jump to $(c_{s}/c)^2 > 1/3$ for densities after the phase transition, preventing the EOS from a corresponding strong stiffening.

\section{Inferring EOS and mass-radius properties}

\subsection{Framework for Bayesian inference}
\label{sec:bayesianmethod}

Next, we describe the statistical framework for constraining the EOS using Bayesian inference, following the protocol outlined in \citet{Riley18}. Using Bayes' theorem, we can write the posterior distribution on the parameters of interest $\bm{\theta}$, in our case the EOS parameters and central densities (interior parameters), as being proportional to a prior distribution $\pi$ times the likelihood $\mathcal{L}$,
\begin{equation}
\label{eq:eq1}
\begin{split}
\mathcal{P}(\bm{\theta} | \mathcal{D}, \mathcal{M}, \mathcal{I}) & = \frac{\pi (\bm{\theta} | \mathcal{M}, \mathcal{I}) \, \mathcal{L} (\mathcal{D} | \bm{\theta}, \mathcal{M})}{\mathcal{P}(\mathcal{D} | \mathcal{M}, \mathcal{I})} \\
& \propto \pi (\bm{\theta} | \mathcal{M}, \mathcal{I}) \, \mathcal{L} (\mathcal{D} | \bm{\theta}, \mathcal{M}) \,.
\end{split}
\end{equation}
where $\mathcal{D}$ denotes an observational dataset, $\mathcal{M}$ the model used, and $\mathcal{I}$ the Bayesian prior information, such as information from previously analyzed datasets. Because the EOS parameters and the central densities are deterministically related to the mass and radius of a neutron star through the relativistic stellar structure equations (the Tolman-Oppenheimer-Volkoff equations in the non-rotating limit), the following must be true for the likelihood: 
\begin{equation}
\label{eq:eq2}
\mathcal{L} (\mathcal{D} | \bm{\theta}, \mathcal{M}) \equiv \mathcal{L} (\mathcal{D} | \bm{M}, \bm{R}, \mathcal{M}) \,.
\end{equation}
Furthermore, for reasons of computational feasibility, we assume
\begin{equation}
\label{eq:eq3}
\mathcal{L} (\mathcal{D} | \bm{M}, \bm{R}, \mathcal{M}) \propto \mathcal{P} (\bm{M}, \bm{R} | \mathcal{D}, \mathcal{M}, \mathcal{I}) \,.
\end{equation}
This follows the approach outlined in Section 2.3.4 of \citet{Riley18}, termed the {\it Interior Prior} paradigm (more robust than the alternative {\it Exterior Prior} method), but uses the approximative marginal likelihood function of the exterior parameters (mass and radius)\footnote{Note that these are not actually computed in this paper, but directly presented as bivariate Gaussian distributions.}, to calculate the marginal posterior function of interior (EOS) parameters. It is a less computationally intensive alternative to full direct inference of EOS parameters from the data. As outlined in \citet{Riley18}, this assumption only holds when the prior on mass and radius, which is implicitly defined in the proportionality, is sufficiently non-informative\footnote{This is expected to be the case for NICER analysis, even for sources like primary target PSR J$0437-4715$ where the well constrained mass arising from radio observations \citep{Reardon16} is treated as a prior; this is because the original radio analysis used a non-informative prior in their computations.}. A second assumption is that the datasets of different observed neutron stars are independent, which allows us to separate the likelihoods and rewrite Eq.~(\ref{eq:eq1}), using Eqs.~(\ref{eq:eq2}) and (\ref{eq:eq3}), as
\begin{equation}
\label{eq:eq4}
\mathcal{P}(\bm{\theta} ~|~ \mathcal{D}, \mathcal{M}, \mathcal{I}) \propto \pi (\bm{\theta}~|~\mathcal{M}, \mathcal{I})~ \prod_{i=1}^s \mathcal{P} (M_i, R_i ~|~ \mathcal{D}_i, \mathcal{M}, \mathcal{I}) \, ,
\end{equation}
for $s$ number of observed stars. This method is numerically similar to the methods used in \citet{Steiner10}, \citet{Ozel16}, and \cite{Raithel17}.

\begin{figure*}
\centering
\includegraphics[width=0.9\textwidth]{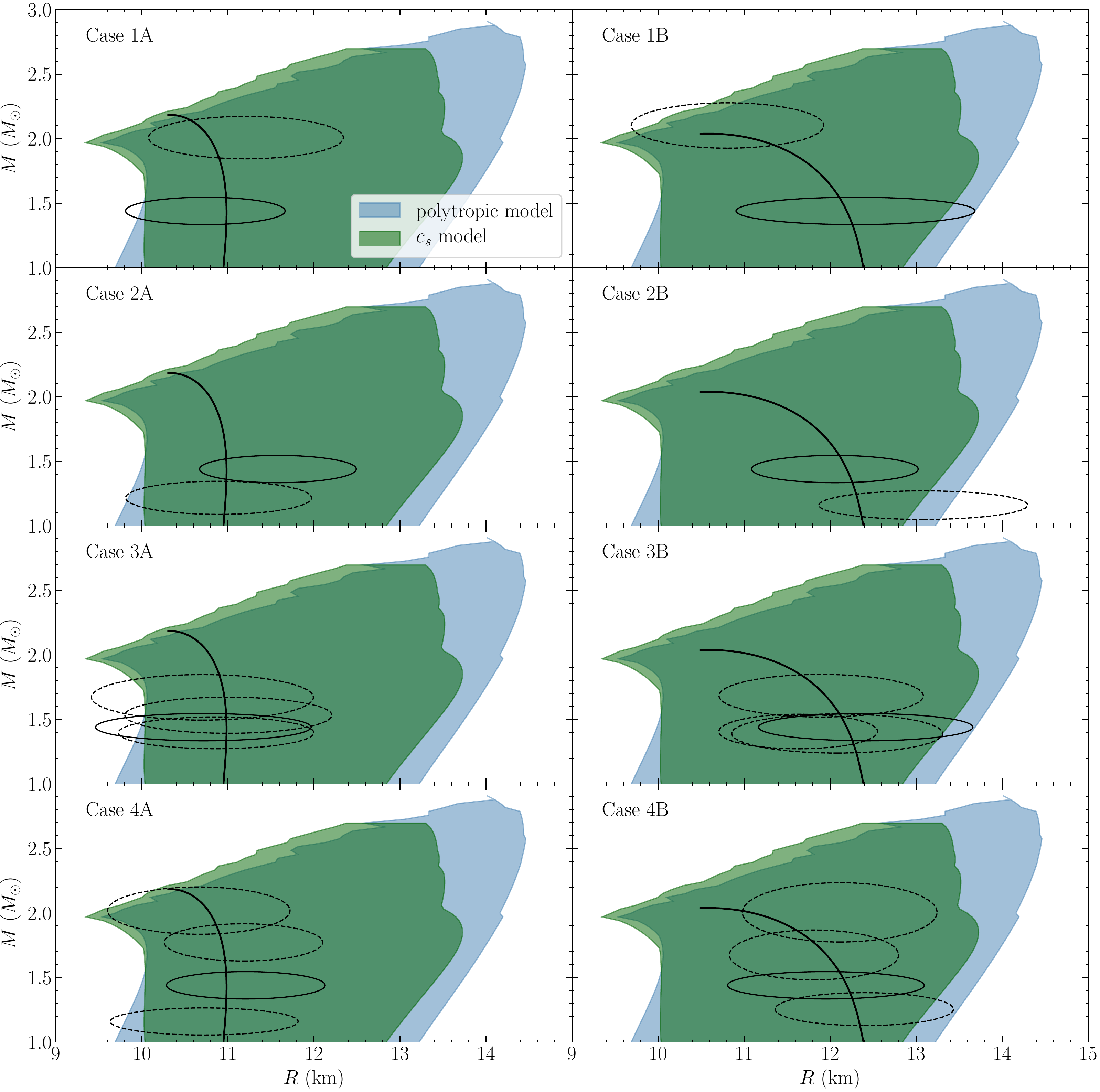}
\caption{Different scenarios of mass and radius posteriors explored in this work. The elliptical contours show the 1$\sigma$ levels of the distributions. The solid contours represent the pulsar PSR J$0437-4715$ with a known mass of $1.44 \pm 0.07 \, M_{\odot}$; the dashed contours are the stars whose mass is not known a priori. The black mass-radius curves correspond to the underlying EOS on which the peak of the Gaussian posteriors are centered before a random scatter was added. The green and blue shaded areas are the full bands given by the CS and PP models, respectively, as in Fig.~\ref{fig:fig2}).}
\label{fig:fig4}
\end{figure*}

\subsubsection{Choice of priors}
\label{sec:priors}

The choice of the prior in Eq. (\ref{eq:eq4}) can play an important role in the inference of EOS parameters \citep{Steiner16} and as such has to be carefully considered. For the PP parametrization as well as for the CS parametrization we use a uniform, continuous prior for all parameters, within the ranges described in Section~\ref{sec:parameterspace}. We impose the five requirements described for the CS parametrization and adopt the requirements from \cite{Hebeler13} for the PP parametrization. The prior on the central energy density of the star is chosen as a uniform prior on $\log(\varepsilon_c)$, with a lower bound of $\log(\varepsilon_c/\mathrm{g}\,\mathrm{cm}^{-3}) = 14.6$ and an upper bound that corresponds to the maximum central energy density reached in a neutron star for that given EOS.

To understand the significance of the prior, we sample its distribution for both models and transform it to the space of pressure and energy density as well as to mass and radius. The resulting prior probability distributions are shown in Fig.~\ref{fig:fig3}, where each histogram contains several times $10^5$ samples. Comparing these distributions to the general bands highlighted in Fig.~\ref{fig:fig2}, one clearly sees much more structure in the distributions than one might naively expect from the bands. For the CS parametrization the structures are qualitatively similar with the sound-speed based parametrization used in \citet{Tews18}. In the space of pressure and energy density both models show a narrow region where the distribution is peaked, with the probability density at a given energy density quickly falling off when moving to higher and lower pressures. For both models these regions encompass reasonably stiff EOS, a consequence of enforcing that the EOS supports a $1.97 \, M_{\odot}$ neutron star. 

The distribution in mass-radius space shows similar structures, with the 68\% credible regions enclosing remarkably narrow radius regions, e.g., for typical $1.4 \, M_{\odot}$ neutron stars less than 1~km for the CS parametrization. From Fig. \ref{fig:fig3} it is also evident that the PP model is even more peaked towards larger radii, especially at masses above $\sim 1.5 \, M_{\odot}$. The apparent bimodality of the 68\% credible regions in both models is a consequence of matching the models to the lower and upper limit of the cEFT band at the transition density $n_{\mathrm{cEFT}}$. The CS model further shows a significant peak just above $2 \, M_{\odot}$. This is a result of the speed of sound decreasing for most EOS at densities around $2 \times 10^{15}$~g$/$cm$^3$ or higher, causing their corresponding mass-radius curves to show only small changes in mass but large changes in radius. The fact that this occurs visibly just above $2 \ M_{\odot}$ is because EOS that do not reach this mass are discarded.

\subsubsection{Numerical methods}

To sample the posterior distribution in Eq. (\ref{eq:eq4}) we use the Python implementation of the Bayesian inference tool \texttt{MultiNest} \citep{Feroz08, Feroz09, Feroz13, Buchner14}. \texttt{MultiNest} makes use of a sampling technique called Nested Sampling \citep{Skilling04}, where a fixed number of parameter vectors is kept throughout the sampling (so-called live points), sorted by their likelihood values and drawn randomly from the prior distribution. The parameter vector with the smallest likelihood is replaced each time with a parameter vector with a higher likelihood, thereby scanning over the full parameter space until the remaining parameter volume becomes small enough and the algorithm terminates. 

The prior in the \texttt{MultiNest} software is always uniformly drawn from the unit hypercube, and thus requires a transformation to comply with a chosen prior. Mostly we want to sample uniformly between prior bounds, which can be easily expressed as 
\begin{equation}
\theta = \theta_{\text{min}} + (\theta_{\text{max}} - \theta_{\text{min}})x \,,
\end{equation}
where $x$ is drawn from the uniform distribution between $0$ and $1$. However, the transition densities in the PP parametrization, $n_{12}$ and $n_{23}$ are subject to the additional requirement that $n_{12} < n_{23}$. To uniformly draw from the triangle $\theta_{\text{min}} < \theta_1 < \theta_2 < \theta_{\text{max}}$ we employ the transformation from \citet{Handley15}\footnote{Note the typo in Eq.~(A13) in \citet{Handley15}, where $x^{1/(n - i +1)}_i$ should be $(1 - x^{1/(n - i +1)}_i$).} 
\begin{equation}
\begin{aligned}
\theta_1 &= \theta_{\text{min}} + (\theta_{\text{max}} - \theta_{\text{min}}) (1 - \sqrt{x}) \,, \\
\theta_2 &= \theta_1 + (\theta_{\text{max}} - \theta_1) (1 - x) \,.
\end{aligned}
\end{equation}
We also note that the prior on central densities is dependent on the other EOS parameters, i.e., $\pi (\bm{\varepsilon_c} | \bm{\theta}, \mathcal{M}, \mathcal{I}) $, which requires the calculation of the full corresponding mass-radius curve to determine the central density where unstable solutions appear. 

The posterior distributions presented in this paper are furthermore calculated with a sampling efficiency of $0.8$ and $5000$ live points to ensure reasonable runtimes. We have made sure, however, that the obtained posteriors are robust to changes in these settings.

\begin{figure*}
\centering
\includegraphics[width=0.9\textwidth]{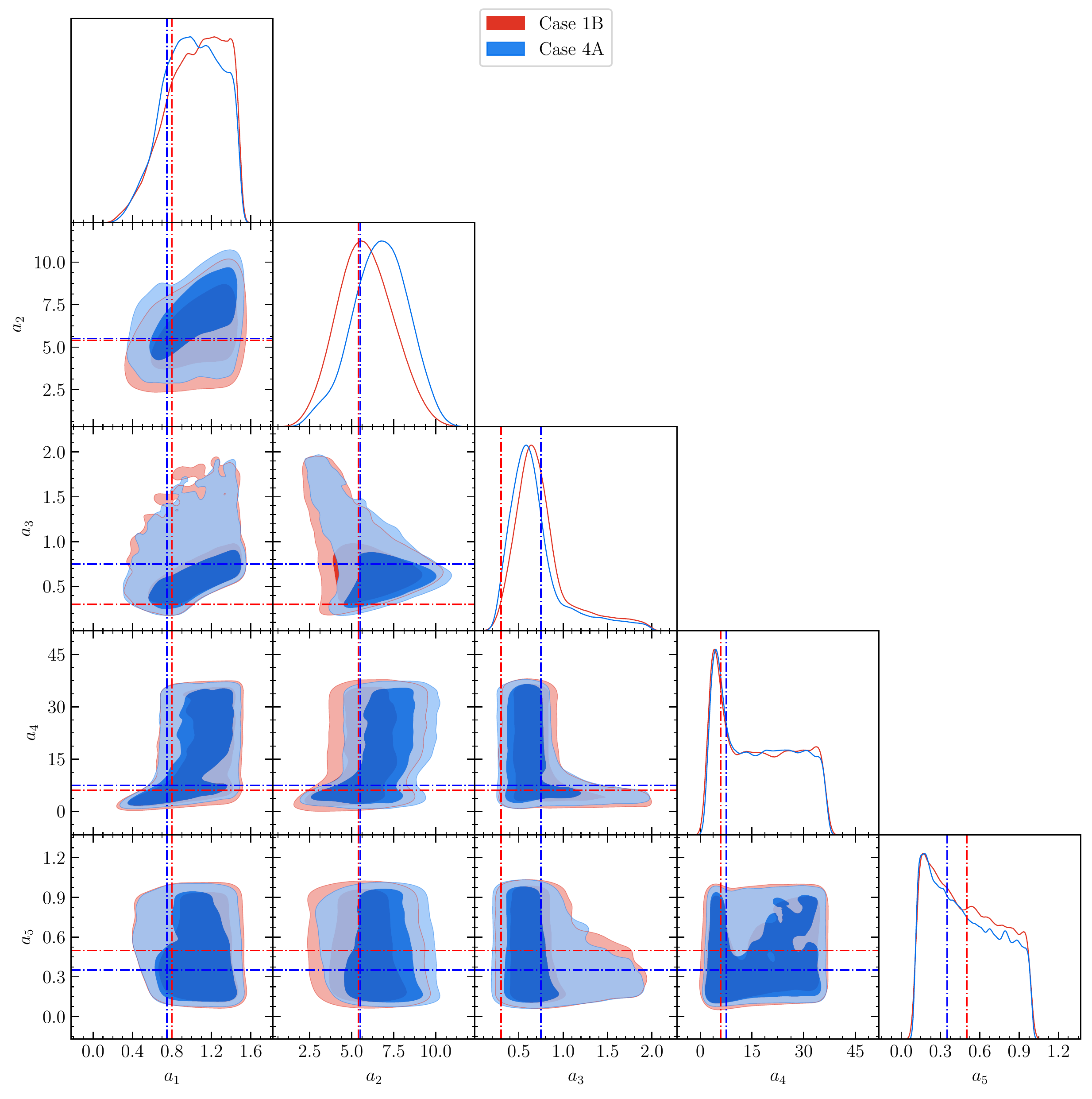}
\caption{Posterior distributions for the EOS parameters for the CS parametrization for Cases 1B (red) and 4A (blue) (see Fig.~\ref{fig:fig4} for both scenarios). The light and dark shaded regions indicate respectively the $68 \%$ and $95 \%$ credible regions of the two-dimensional marginalized posteriors.  The line plots are the one-dimensional marginalized posteriors for each parameter. The dash-dotted blue and red lines give the parameters describing the underlying EOS used to generate the mass-radius posteriors (see Section~\ref{sec:configurations}).}
\label{fig:fig5}
\end{figure*}

\subsubsection{Bayes factors}
\label{sec:bayesfactors}

To make a quantitative comparison between the CS and PP models we explore their posterior odds \citep{Jeffreys98}, which is defined as the ratio between their posterior probabilities. Using Bayes' theorem we can write this as:
\begin{equation}
\frac{\mathcal{P}(\mathcal{M}_{\textrm{cs}} | \mathcal{D}, \mathcal{I})} {\mathcal{P}(\mathcal{M}_{\textrm{pp}}  | \mathcal{D}, \mathcal{I})}  = B ~ \frac{\mathcal{P}(\mathcal{M}_{\textrm{cs}} | \mathcal{I})}{\mathcal{P}(\mathcal{M}_{\textrm{cs}} | \mathcal{I})} = \textrm{Bayes factor} \times \textrm{prior odds} \,,
\end{equation}
where the Bayes factor is defined as the ratio between what is called the evidence or sometimes the marginal likelihood:
\begin{equation}
B = \frac{\mathcal{P}(\mathcal{D} | \mathcal{M}_{\textrm{cs}}, \mathcal{I})}{\mathcal{P}(\mathcal{D} | \mathcal{M}_{\textrm{pp}}, \mathcal{I})}.
\end{equation}
If we assume that both models are equally probable \textit{a priori}, we can use the Bayes factor to compare between the two models. The calculation of these factors is straightforward given that \texttt{MultiNest} automatically computes the evidence for each model.

\subsection{Configurations of mass-radius posterior distributions}
\label{sec:configurations}

In order to compare different methods of constraining the EOS and the effect the parametrization has on these constraints we explore multiple scenarios of mass-radius posterior distributions. All distributions are modeled as bivariate Gaussian distributions
\begin{equation}
\mathcal{P}(M, R | \mathcal{D}, \mathcal{M}, \mathcal{I}) = \frac{\sigma_M \sigma_R}{2} \exp \left[ - \frac{(M - \mu_M)^2}{2\sigma_M^2} - \frac{(R - \mu_R)^2}{2\sigma_R^2} \right] \,,
\end{equation}
with the mean of the distribution centered on a specific underlying EOS. Note that realistic mass-radius posteriors expected from the waveform modelling technique used by NICER will have some degeneracy between mass and radius \citep[see, e.g.,][]{Miller15}. However, the differences that might result from different parametrizations of the EOS can be illustrated using simplified posteriors, without a mass-radius degeneracy. 

For each scenario of different mass sources, we consider two different underlying EOS: a relatively soft, standard EOS with a radius around 11~km (labelled A); and a more extreme EOS covering a larger spread in radii (labelled B). We then define scenarios that may emerge as a result of the NICER observations.

For the first two scenarios we consider the two primary science targets of NICER \citep{NICER}: the pulsar PSR J$0437-4715$ with a mass of $1.44 \pm 0.07 \, M_{\odot}$ \citep{Reardon16} and PSR J$0030+0451$, for which the mass is unknown. In Case 1 we assume that the mass of this pulsar is $2.0$ M$_{\odot}$ and in Case 2 that it is $1.2$ M$_{\odot}$. In Case 3 and 4 we add two more stars, so that we have four mass-radius posteriors. This is representative of the results eventually expected from NICER. The next two highest priority targets being studied by NICER are PSR J1231+1411 and PSR J2124-3358; for neither of these stars the mass is known.  For Case 3 we assume that the three unknown masses lie relatively closely together: $1.4 \, M_{\odot}$, $1.5 \, M_{\odot}$, and $1.7 \, M_{\odot}$. In Case 4 we take them to be more widely spread: $1.2 \, M_{\odot}$, $1.7 \, M_{\odot}$, and $2.0 \, M_{\odot}$. This is obviously far from exhaustive, but lets us explore a range of representative scenarios.

We then add a random scatter to all masses and radii drawn from a Gaussian distribution centered on the EOS with standard deviation of $3 \%$ of the chosen mass and radius values, except for the known neutron star mass. The uncertainties of the distributions, $\sigma_M$ and $\sigma_R$, are randomly picked from a uniform distribution between $5-10 \%$ of the central mass-radius values, except again when the mass is known. As each of these configurations is considered with two different underlying EOS, we have a total number of eight scenarios, depicted in Fig.~\ref{fig:fig4}. 

\begin{figure*}
\centering
\includegraphics[width=0.9\textwidth]{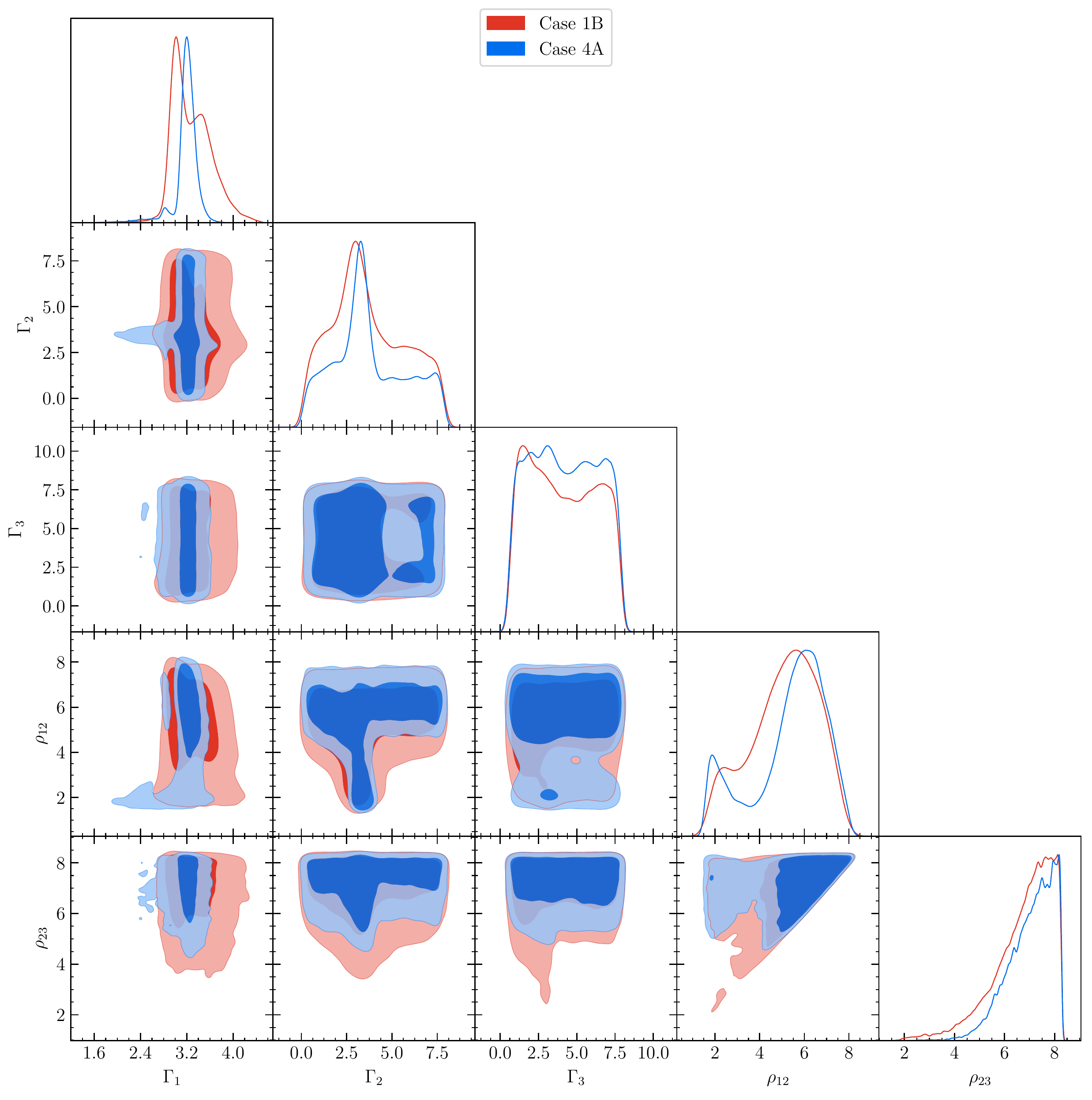}
\caption{Same as Fig.~\ref{fig:fig5} but for the PP parametrization.}
\label{fig:fig6}
\end{figure*}

\subsection{Posterior distributions}
\label{sec:posteriors}

\subsubsection{Interior parameter space}

For each scenario described in the previous section we obtain from \texttt{MultiNest} a set of equally-weighted posterior samples. The posterior distribution on the EOS parameters can then be estimated by binning these samples and applying a smoothing kernel density estimation\footnote{In this case we have used a Gaussian kernel density estimation, which means that each bin is estimated as a Gaussian and weighted by its frequency. The full distribution is then a smooth summation of all the individual Gaussians. To determine the parameter $k$ that controls the smoothing, we have used Scott's Rule \citep{Scott92}, i.e., $k = n^{-1/(d+4)}$, where $n$ is the number of datapoints and $d$ the number of dimensions.}. Two examples (for Cases 1B and 4A) are given in Figs.~\ref{fig:fig5} and~\ref{fig:fig6} for the five EOS parameters in the CS and PP models. In each subplot the distribution is marginalized over the parameters not shown. For the CS model we include the parameters describing the underlying EOS used to generate the mass-radius posteriors. These might not necessarily be the parameters that receive the most support from the likelihood after adding a random scatter, but they still represent an EOS that is consistent with the mass-radius posteriors.

\begin{figure*}
\centering
\includegraphics[width=0.9\textwidth]{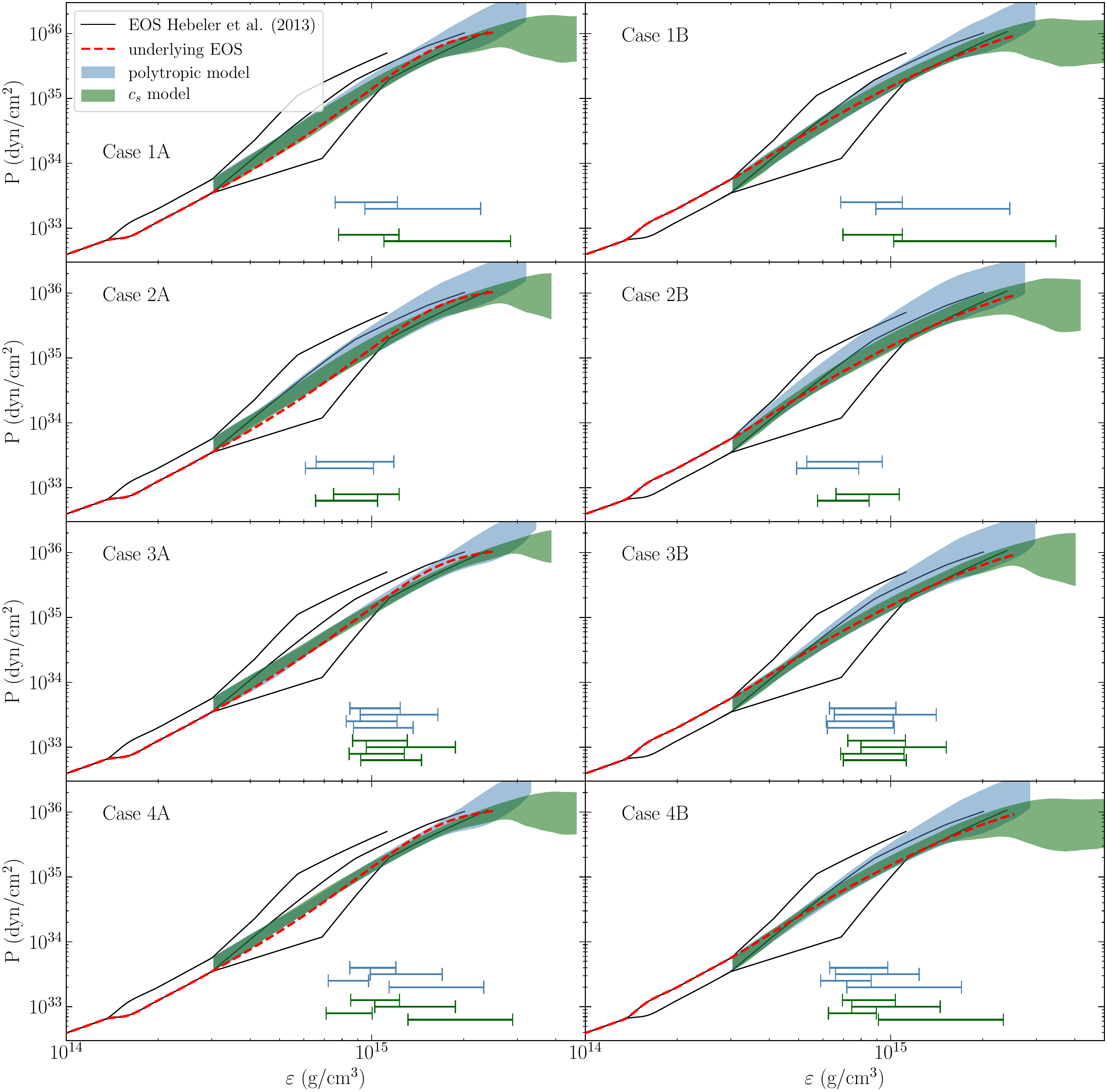}
\caption{$95 \%$ credible regions of the posterior distributions for the EOS for the CS (green) and the PP model (blue). For comparison, the red dashed line represents the underlying EOS used to generate the mass-radius posteriors, while the black EOS are the three representative EOS from \citet{Hebeler13}. The narrow features of the green and blue regions are a result of the prior and the likelihood peaking in different regions of parameter space. Especially for Cases A the posterior distribution follows closely the edge of the priors in Fig.~\ref{fig:fig3}, indicating that the posteriors are not completely likelihood-dominated. The green and blue horizontal bars in each panel give the $95 \%$ confidence interval for the marginalized posterior distribution of the maximal central energy density reached in neutron stars, for the CS and PP model according to the color code.}
\label{fig:fig7}
\end{figure*}

\begin{figure*}
\centering
\includegraphics[width=0.9\textwidth]{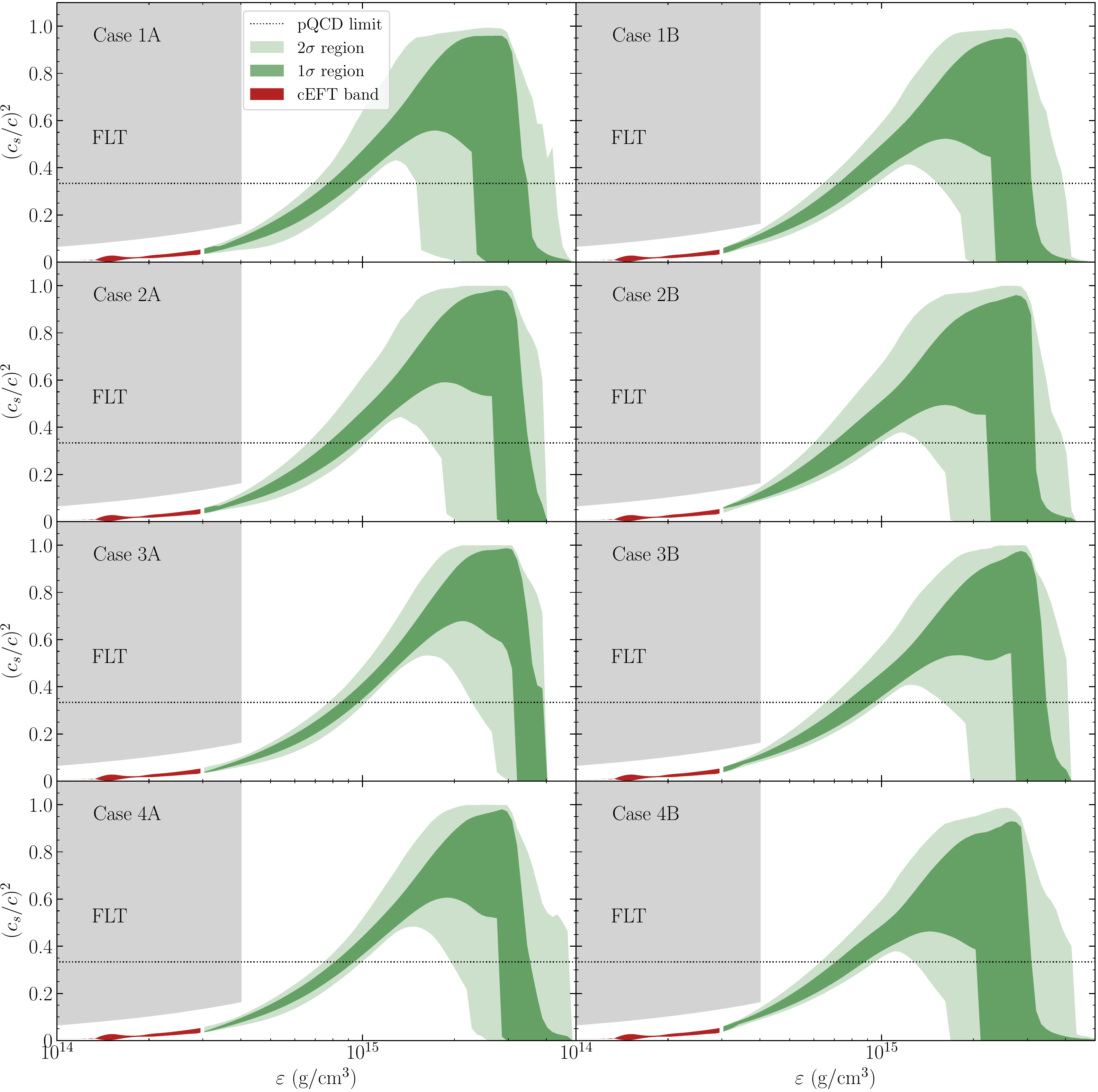}
\caption{Posterior distributions of the speed of sound in the CS model, where the dark and light green regions represent the joined $68 \%$ and $95 \%$  credible regions at discrete energy densities. The red area at lower densities gives the speed of sound of the cEFT band calculated by \citet{Hebeler10}. The dotted line indicates the value $1/\sqrt{3}$ of the speed of sound in the asymptotic pQCD limit, and the grey area is the excluded region by the Fermi liquid theory (FLT) constraints.}
\label{fig:fig8}
\end{figure*}

\begin{figure*}
\centering
\includegraphics[width=0.9\textwidth]{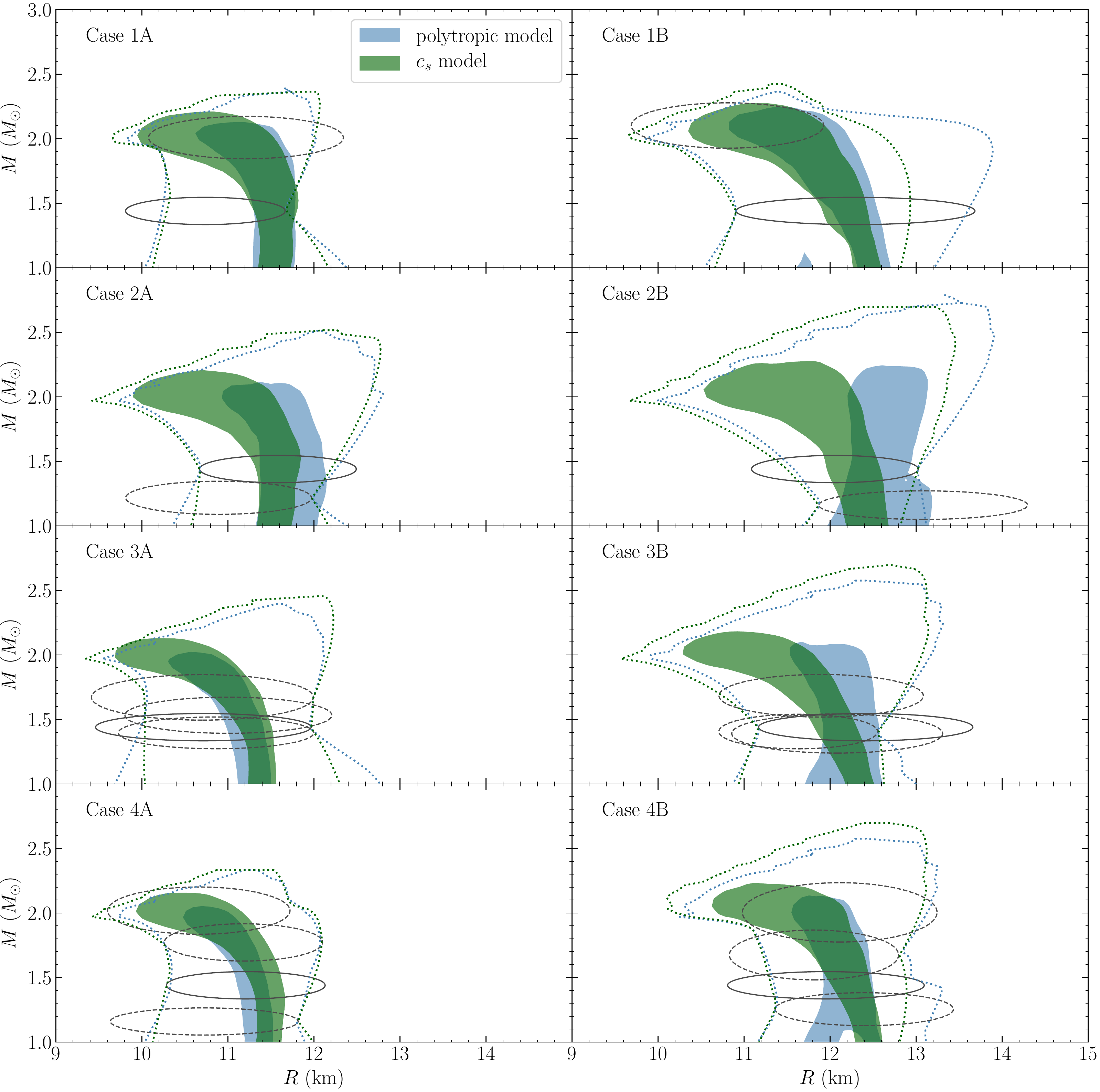}
\caption{68\% credible regions of the posterior distribution for the mass and radius for the CS (green) and the PP (blue) models. The dotted lines in contrast illustrate the band of all EOS that generate mass-radius curves passing through the $1\sigma$ contours of the likelihood. These are much wider than the posterior distributions obtained from the Bayesian inference. The elliptical contours of the input mass-radius posteriors for each scenario are shown as well for reference. As in Fig.~\ref{fig:fig4} the solid contour is for PSR J$0437-4715$ with a known mass of $1.44 \pm 0.07 \, M_{\odot}$, and the dashed ellipses are the other sources.}
\label{fig:fig9}
\end{figure*}

We can translate this posterior distribution to the space of the EOS by discretizing $\varepsilon_{i}$ onto a grid and calculating for each posterior sample the pressure $P = P(\varepsilon_{i})$. From these pressure values we create a set of one-dimensional histograms at an $\varepsilon_i$ and subsequently calculate the $95 \%$ credible region. The individual credible regions at each $\varepsilon_i$ are then joined together to obtain a band that represents the $95 \%$ credible region of the posterior distribution for the EOS. This is shown in Fig.~\ref{fig:fig7}. A striking feature of these bands is the narrowing at intermediate densities, which suggests that tight constraints on the physics of dense matter are possible. \footnote{As a check, we determined the predicted radii of a typical neutron star with a mass of $1.4 \, M_{\odot}$ and the pressure at twice saturation density for each EOS inside the uncertainty band. We found that $R_{1.4 \, M_{\odot}}$ is strongly correlated with $P(2 \, n_0)$, which is consistent with the findings by \citet{Latt13PRcorr}.} In most cases the underlying EOS falls within these bands, but in some A scenarios the underlying EOS lies slightly outside for some energy densities. This is a consequence of the prior constraints, which lead to stiffer EOS receiving more prior support (see Fig.~\ref{fig:fig3} and the discussion in Sect.~\ref{sec:priors}), which is closer to the B scenarios. When the likelihood encompasses softer EOS, as in the A scenarios, the posterior distribution consequently peaks in the region that has finite support from both the prior and the likelihood, so that the posteriors get shifted to stiffer EOS. Moreover, the horizontal bars in each panel of Fig.~\ref{fig:fig7} give the $95 \%$ confidence interval for the marginalized posterior distribution of the maximal central energy density reached in neutron stars. This shows the highest central densities that are relevant to neutron stars, which are well below the asymptotic pQCD regime.

In Fig.~\ref{fig:fig8} we show the corresponding bands for the speed of sound for the CS model. The dark and light green bands correspond to the $95 \%$ and $68 \%$ credible regions, respectively. For the scenarios shown, the constraints from FLT at lower densities have no significant impact on the posterior distributions. The FLT constraints would become important if a large and heavy neutron star were to be included in the mass-radius posterior distributions (see Fig.~\ref{fig:fig4}). With increasing densities the speed of sound increases monotonically well beyond $(c_s/c)^2 = 1/3$ up to energy densities exceeding $10^{15}$~g/cm$^3$ for all considered scenarios. Only close to the maximal central energy density (see horizontal bars in Fig.~\ref{fig:fig7}), when the maximal mass of the neutron star has been reached, does the speed of sound tend to decrease below this value again. This is due to the softening necessary to remain causal. This shows that the pQCD constraints used in the CS parametrization (see Section~\ref{sec:CS_parametrization}) are important only for densities well beyond the regime that is relevant for typical neutron stars.

\subsubsection{Exterior parameter space}

For the parameter space of neutron star masses and radii we show the posterior predictive distribution, which gives the probability of a new mass-radius point given the posterior distribution of the EOS parameters. To avoid redrawing samples from the posterior distributions we use the posterior samples obtained in our analyses, marginalize over central densities and draw a new central density from their prior distribution. Numerically this results in a set of mass-radius points for which we can calculate the $68 \%$ credible region by binning and performing kernel density estimation. 

We show the credible regions for these posterior predictive distributions for all the scenarios considered in Fig.~\ref{fig:fig9}, for both the PP and CS parametrizations. In most cases both parametrizations result in similar bands in mass-radius space, however, there are also significant differences between the two parametrizations. In all cases where the likelihood is centered around lower-mass stars, the PP models allow for a larger region at larger radii, especially in Cases 2B and 3B. This is a direct consequence of the form of the parametrization, as the PP model includes EOS that produce mass-radius curves with almost constant radius up to high masses. The speed of sound model however does not permit these kinds of EOS due to the form of the Gaussian, which forces every EOS to soften again after the peak of the Gaussian to comply with the pQCD constraint. Note that the small bimodal feature for the PP parametrization at low masses in Cases 1B and 4B is a consequence of the way the polytropic extensions are matched to the upper and lower limit of the cEFT band.

In addition to the posterior distributions in Fig.~\ref{fig:fig9}, we also show with dotted lines the region one would obtain when discarding all EOS from the PP and CS band that do not produce mass-radius curves going through all $1\sigma$ contours of the mass-radius posteriors. This could be termed a simple compatibility cut. We note that in general these regions could be used as a very conservative estimate of EOS that would have a finite probability when inference is performed. One has to be careful though: in Case 1A and 2A this region would exclude few EOS that are within the $68 \%$ credible region of the posterior distribution; and in general Bayesian inference of the parameters provides much tighter constraints, which however requires that the prior assumptions and sampling are fully understood.

\begin{figure}
\centering
\includegraphics[width=0.9\columnwidth]{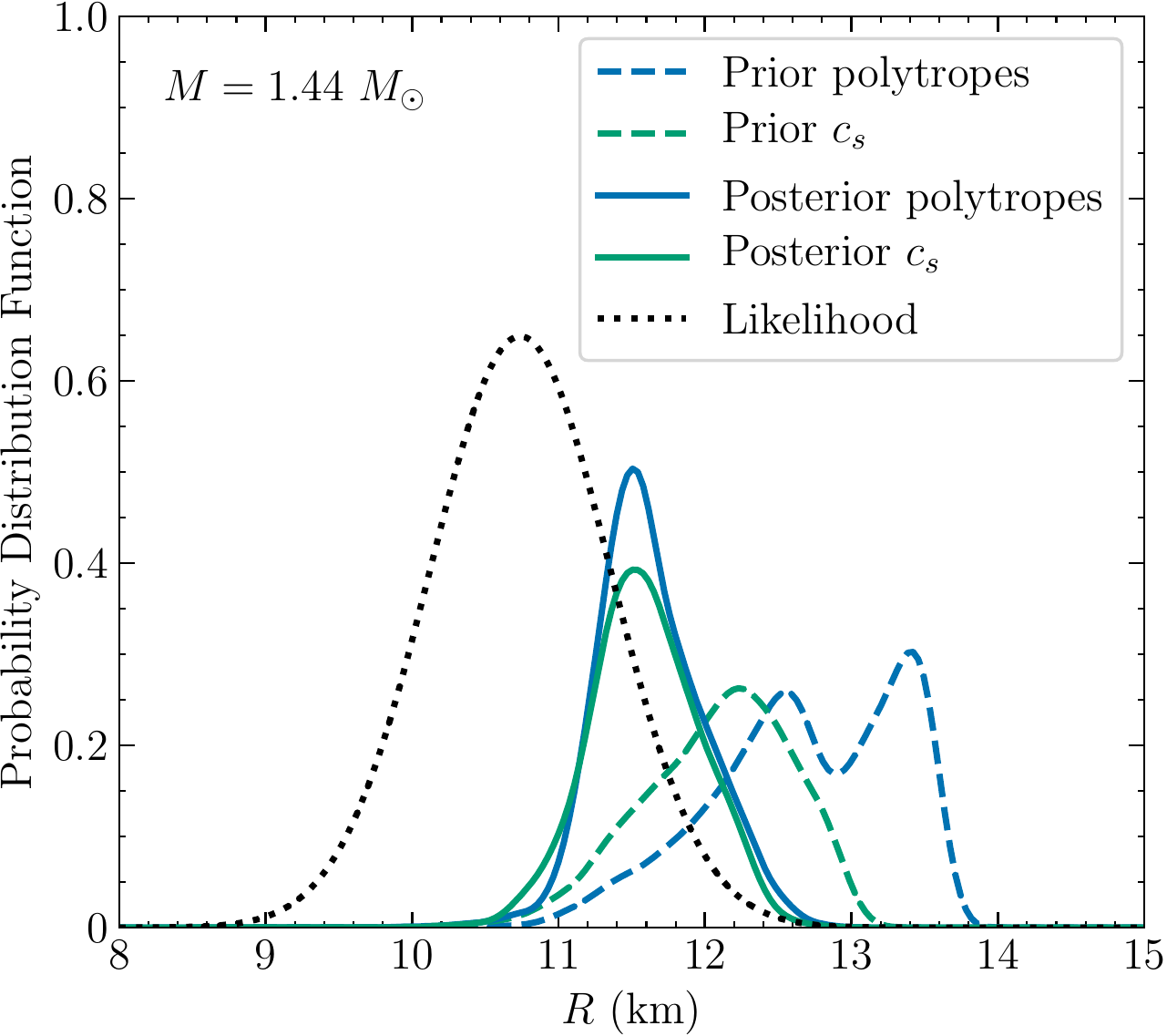}
\caption{One-dimensional cut for a $1.44 \, M_{\odot}$ star of the probability distributions of the priors for both parametrizations and the likelihood given by the ellipse of the $1.44 \, M_{\odot}$ star (PSR J$0437-4715$) of Fig.~\ref{fig:fig3}. In addition, we show the resulting posteriors for both parametrizations, both for Case 1A.}
\label{fig:fig10}
\end{figure}

For all posterior distributions for the A scenarios the $68 \%$ credible regions for both the PP and CS models seem centered towards larger radii than one might expect. This behavior again follows from the prior used on the EOS parameters. To better understand how the uniform prior on EOS parameters affects the posterior distribution, we show in Fig.~\ref{fig:fig10} a one-dimensional cut for a $1.44 \, M_{\odot}$ star of the probability distributions of the priors for both parametrizations and the likelihood given by the ellipse of the $1.44 \, M_{\odot}$ star (PSR J$0437-4715$) of Fig.~\ref{fig:fig3}. Figure~\ref{fig:fig10} illustrates clearly that the posterior distribution is not completely likelihood-dominated, due to the prior pushing towards larger radii. As a result, there is only a small region of parameter space around 11.5~km where there is both finite support from the likelihood and the prior, leading to an unexpectedly peaked posterior for the radius and the narror regions for the mass-radius bands in Fig.~\ref{fig:fig10}. 

\subsubsection{Bayes factors}

In order to compare the CS and PP model quantitatively we use the evidences computed by \texttt{MultiNest} to calculate the Bayes factors (see Section \ref{sec:bayesfactors}). The values are reported in Table \ref{table:tab1}, where $B \geq 1$ and $B \leq 1$ indicates more support for the PP model and the CS model, respectively. Following the interpretation provided by \citet{Kass95} none of the Bayes factors shown here indicate that one of the two models is more favoured by the data.

\begin{table}
\centering
\begin{tabular}{@{}lll@{}}
\toprule
Scenario & \multicolumn{2}{c}{Bayes factor }     \\ \midrule
 & \multicolumn{1}{c}{A} & \multicolumn{1}{c}{B} \\ \midrule
Case 1 & 0.653   $\pm$ 0.026 & 0.416   $\pm$ 0.016\\
Case 2 & 1.320   $\pm$ 0.055 & 1.962   $\pm$ 0.079 \\
Case 3 & 1.615   $\pm$  0.100 &  1.433   $\pm$  0.083 \\
Case 4 & 0.821   $\pm$ 0.050 & 2.012   $\pm$  0.111 \\ \bottomrule
\end{tabular}
\caption{The computed Bayes factors for comparing the PP model to the CS model.}
\label{table:tab1}
\end{table}

\section{Discussion and conclusions}

In this work, we have explored constraints on the EOS of dense matter resulting from future measurements of neutron star masses and radii, combined with EOS constraints from nuclear physics, a two-solar-mass neutron star, and causality. To this end, we employed a Bayesian inference framework and considered different scenarios of neutron star observations that reflect possible outcomes of the ongoing NICER mission. By using two different EOS parametrizations we demonstrated how constraints on properties of dense matter and neutron stars can be inferred from such measurements and how to probe the sensitivity of the results to particular descriptions of the EOS.

In addition to the well-established PP parametrization, we have developed an alternative description based on the speed of sound in a neutron star. We argue that such a parametrization makes more physical connections than the PP parametrization, as it produces EOS with a continuous speed of sound, can take into account constraints based on Fermi liquid theory, and complies with the asymptotic high-density limit from pQCD calculations, although the latter are well beyond the densities reached in neutron stars. In Section~\ref{sec:parameterspace} we showed that the PP parametrization used in \citet{Hebeler13} and the introduced CS model generate a relatively similar band of EOS after incorporating the same constraints from nuclear physics at lower densities, a two-solar-mass neutron star, and causality.

Using these two EOS parametrizations, we have performed parameter estimation for eight different scenarios of possible posterior mass-radius distributions, either with $2$ or $4$ stars. We find that the difference in the resulting posterior distributions between the two EOS parametrizations is most notable when all stars have low masses, as the CS model produces fewer EOS with an almost constant radius. For the scenarios where the stars have smaller radii, posterior distributions of the CS model are more peaked towards softer EOS, although the posterior for both models seems shifted towards larger radii than one would expect from the likelihood function.   

The offset between the mass-radius likelihood distributions and the inferred posterior distributions for soft EOS scenarios is a consequence of the uniform prior on the EOS parameters and other prior constraintumed in all scenarios. In Fig.~\ref{fig:fig3} the prior in the EOS parameters is shown to map to a prior on mass and radius that peaks towards larger radii. As a result the posterior distributions are not completely likelihood dominated, although still in good agreement with the $1\sigma$ regions of the mass-radius posteriors. 

%It might be possible to choose a prior that results in a more uniform distribution when translated to mass and radius, but the distribution in Fig.~\ref{fig:fig3} is a consequence of reasonable assumptions regarding the nature of dense matter around saturation density, together with the observational constraint of the measured pulsar PSR J$0348+0432$ \citep{Antoniadis13} with a mass of $2.01 \pm 0.04 \, M_{\odot}$. There is, however, still some freedom in the way the parameter space is sampled. Further research is then required to investigate whether, e.g., parameters drawn logarithmically result in more uniform distributions in EOS space and mass-radius space. Furthermore, if in the future neutron stars with low radii were measured, or with lower uncertainty, then the posterior distribution would be more likelihood dominated, shifting the peak of the distribution to smaller radii as well. 

It might be possible to choose a prior that results in a more uniform distribution in either EOS space or when translated to mass-radius space (due to the complex mapping of the TOV equations it is not clear that one can achieve both simultaneously), but the distribution in Fig.~\ref{fig:fig3} is a consequence of reasonable assumptions regarding the nature of dense matter around saturation density, together with the observational constraint of the measured pulsar PSR J$0348+0432$ \citep{Antoniadis13} with a mass of $2.01 \pm 0.04 \, M_{\odot}$. There is, however, still some freedom in the way the parameter space is sampled. Further research is then required to investigate whether, e.g., parameters drawn logarithmically result in more uniform distributions in EOS space and mass-radius space. One alternative might be to use Gaussian processes to generate a non-parameteric EOS, as in the recent paper by \citet{Landry18} on EOS inference from gravitational wave measurements; however even in that case posteriors were found to be prior-dominated due to limited data. Another alternative would be to ensure that any peaking in the distribution is genuinely physically motivated, rather than a somewhat inadvertent consequence of trying to cover a range of parameter space, as it is for the models examined in this paper. If in the future neutron stars with low radii were measured, or with lower uncertainty, then the posterior distribution would be more likelihood dominated, shifting the peak of the distribution to smaller radii as well. In principle the approach that we have outlined in this paper can be used to determine the level of uncertainty on $M-R$ posteriors necessary to ensure that EOS measurements are likelihood dominated.  A comprehensive answer to this question will depend on the precise spread of masses and the exact shape of the $M-R$ posteriors (which could be multimodal).  One should also consider the impact for a broader range of input EOS, and the fact that it may only be necessary to reduce the uncertainty (by increasing observing time) on a subset of the $M-R$ posteriors. Once preliminary NICER results are available, it would be extremely valuable to carry out such an analysis.

Due to this sensitivity of the results to the prior distribution the interpretation of the posterior distributions as shown in Figs.~\ref{fig:fig7}--\ref{fig:fig9} requires some care. In particular, it will be key to systematically study the effects of different choices for sampling the individual EOS before robust conclusions on constraints for the EOS and neutron star radii can be drawn. For example, if one naively discards all EOS that are not within the $95 \%$ credible region of the posterior distribution of Case 2A or Case 4A, the underlying EOS would not be recovered (see Fig.~\ref{fig:fig9}).

It could be argued that given the prior distribution on the EOS in Fig.~\ref{fig:fig3}, our choice of a relatively soft EOS to center the mass-radius posteriors on is not reasonable. However, inference using X-ray spectral modeling of neutron stars seems to favour radii in the range of $9 - 13$~km \citep[see, e.g.,][]{Steiner13, Bogdanov16, Ozel16,Nattila16,Nattila17,Shaw18}, although the methods are heavily affected by systematics that are still to be resolved \citep[see, e.g.,][]{Watts16} and the sensitivies revealed in this work also need to be accounted for. Another constraint on the neutron star radius comes from the tidal deformability effects on the waveform of a binary neutron star inspiral, first detected in August 2017 \citep{Abbott17}. The $90 \%$ credible region of the posterior distribution on radii derived from this event falls roughly in the range of $10 - 14$~km \citep{Annala18,Abbott18, Most18,Tews18b,Lim18tidaldef}. The soft EOS chosen in this paper, with a radius of $~11$ km, is therefore well within the range of radii quoted in the literature. We stress that the presented EOS and mass-radius regions in this paper are not based on real observational data and hence cannot be directly compared to the extractions from these references. However, our results demonstrate the significance of the EOS priors and other EOS sensitivities in the inference, which suggests that some of these analyses may have to be revisited for a full statistical interpretation of the inferred EOS and neutron star properties.

\section*{Acknowledgements}

This is a joint lead author paper: SG led the CS parametrization development and PP calculations; GR led the inference computations. The work of SG, KH and AS is funded by the Deutsche Forschungsgemeinschaft (DFG, German Research Foundation) -- Projektnummer 279384907 -- SFB 1245. GR and ALW acknowledge support from ERC Starting Grant No. 639217 CSINEUTRONSTAR (PI Watts), and would like to thank Thomas Riley for helpful discussions. This work has benefited from the collaborative network established and supported by the National Science Foundation under Grant No. PHY-1430152 (JINA Center for the Evolution of the Elements).

\bibliographystyle{mnras}
\bibliography{References}

% Don't change these lines
\bsp	% typesetting comment
\label{lastpage}
\end{document}